\documentclass[smallcondensed]{svjour3}       
\smartqed  
\usepackage{amsmath}
\usepackage{graphicx}
\usepackage[utf8]{inputenc}

\usepackage[hyphens]{url} 
\usepackage{hyperref}

\usepackage{color}
\usepackage{fancyvrb}

\DefineVerbatimEnvironment{Highlighting}{Verbatim}{commandchars=\\\{\}}
\usepackage{framed}
\definecolor{shadecolor}{RGB}{248,248,248}
\newenvironment{Shaded}{\begin{snugshade}}{\end{snugshade}}

\newcommand{\CommentTok}[1]{\textcolor[rgb]{0.56,0.35,0.01}{\textit{#1}}}

\newcommand{\KeywordTok}[1]{\textcolor[rgb]{0.13,0.29,0.53}{\textbf{#1}}}
\newcommand{\NormalTok}[1]{#1}
\newcommand{\OperatorTok}[1]{\textcolor[rgb]{0.81,0.36,0.00}{\textbf{#1}}}

\newcommand{\StringTok}[1]{\textcolor[rgb]{0.31,0.60,0.02}{#1}}

\usepackage{booktabs}
\usepackage{longtable}
\usepackage{array}
\usepackage{multirow}
\usepackage{wrapfig}
\usepackage{float}
\usepackage{colortbl}
\usepackage{pdflscape}
\usepackage{tabu}
\usepackage{threeparttable}
\usepackage{threeparttablex}
\usepackage[normalem]{ulem}
\usepackage{makecell}
\usepackage[table]{xcolor}
\usepackage{natbib}

\begin{document}

\title{Simulation-Based Decision Making in the NFL using NFLSimulatoR \thanks{Correspondence should be addressed to Benjamin Williams.} }

    \titlerunning{Decision Making in the NFL}

\author{Benjamin Williams \and  Will Palmquist \and  Ryan Elmore}

    \authorrunning{ Williams, Palmquist, and Elmore }

\institute{
        Benjamin Williams \at
     Department of Business Information and Analytics, Daniels College of
 Business, University of Denver \\
     \email{\href{mailto:Benjamin.Williams@du.edu}{\nolinkurl{Benjamin.Williams@du.edu}}}  
    \and
        Will Palmquist \at
     Department of Business Information and Analytics, Daniels College of
 Business, University of Denver \\
     \email{\href{mailto:Will.Palmquist@du.edu}{\nolinkurl{Will.Palmquist@du.edu}}}  
    \and
        Ryan Elmore \at
     Department of Business Information and Analytics, Daniels College of
 Business, University of Denver \\
     \email{\href{mailto:Ryan.Elmore@du.edu}{\nolinkurl{Ryan.Elmore@du.edu}}}  
    }

\date{Received: date / Accepted: date}

\maketitle

\begin{abstract}
In this paper, we introduce an R software package for simulating plays
and drives using play-by-play data from the National Football League. The simulations are generated by sampling play-by-play data from previous football seasons. The sampling procedure adds statistical rigor to any decisions or inferences arising from examining the simulations. We highlight that the package is particularly useful as a data-driven tool for evaluating potential in-game strategies or rule changes within the league.
We demonstrate its utility by evaluating the oft-debated strategy of
``going for it'' on fourth down and investigating whether or not teams should pass more than the current standard.
\\

\end{abstract}
\clearpage

\def\spacingset#1{\renewcommand{\baselinestretch}%
{#1}\small\normalsize} \spacingset{1}


\hypertarget{intro}{%
\section{Introduction}\label{intro}}

Data-driven decision making is a ubiquitous strategy in today's
marketplace and is becoming increasingly common amongst professional
sports organizations. From the Major League Baseball's Moneyball
movement \citep{lewis03} to the Moreyball strategies employed by the National Basketball Association's Houston Rockets \citep{walsh19},
analytics are no longer the sole purview of the academy as teams try to
improve their performance by investigating the data. The general
consensus, however, is that the National Football League (NFL) lags
behind other professional sports leagues in their use of analytics
\citep{clark}. This does seem to be changing, as evidenced by a recent
hiring trend of data analysts to NFL teams \citep{loque19}, as well as
within the league office in New York.

In the NFL, perhaps the most widely debated research question regards
the decision of going for it on fourth down. This has also been the
subject of several academic articles \citep{romer2006firms,yam2019lost},
the New York Times ``4th down bot'' \citep{fdbot, causey15}, and a new calculator from sports analyst Ben Baldwin \citep{baldwin_athletic}. The consensus among researchers and analysts is NFL coaches tend to be too conservative in their fourth-down calls, often preferring to kick the football (punt or field goal attempt) when the data suggests they should pass or run the ball.

While the decision to go for it on fourth down is much discussed, there
are a plethora of other strategies a team may wish to investigate.
Potential strategies for deeper investigation range from the frequency
and type of plays run, the use of a team's (limited) timeouts during a
game, defensive alignment, and so on. The seemingly infinite
possibilities for NFL strategy evaluation made us wonder how one could
determine which strategies offer the best chance of winning. Some
attempt has been done for strategies such as passing versus running the
football. In the sole peer-reviewed article that we are aware of, \cite{levitt09} found NFL teams did not pass as much as they should.
\cite{hermsmeyer} similarly noted, in an article for the data journalism
website fivethirtyeight.com, that even though the NFL has transitioned
to become a more passing heavy league, teams should still pass more.
Apart from the passing versus rushing and fourth down decision making,
there is a lack of research regarding NFL strategies in the literature.

In this paper we present an R software package, \texttt{NFLSimulatoR},
and an analytically rigorous method for analyzing NFL strategies. Our
method consists of simulating strategies via the sampling of NFL
play-by-play datasets realized in previous seasons. This
simulation method is flexible and allows for the investigation of many
possible strategies and offers a tool for informed decision making with
respect to sport performance. We have embedded the simulation framework
into an open source software package to share the method with the
broader sports analytics community. The rest of the paper is outlined as
follows. In the next section we present the R software package we wrote
for simulating NFL strategies. Section 3 describes the use of the
software package for the two strategies we have discussed thus far:
fourth down decision making and passing versus rushing. Finally, we
offer some concluding thoughts about using (and contributing to) the
package moving forward and other discussions in the final section.

\hypertarget{sec:nfl_sim}{%
\section{NFLSimulatoR}\label{sec:nfl_sim}}

The ideas presented in this paper are, in part, inspired by a blog post
by Mike Lopez, currently the Director of Data and Analytics for the NFL,
in which he used a simulation-based approach to investigate a potential
overtime rule change in the NFL \citep{statsbylopez}. In contrast to the one-off solution presented on his blog, we provide a robust software platform for assessing NFL strategies in the \texttt{NFLSimulatoR} R package. Our desire is for the wider analytics community to use this package, extend our work, and study other strategies in a analytically sound manner.

The ideas embedded in \texttt{NFLSimulatoR} are simple, yet extremely
powerful. The key feature is that we rely on simulations of actual NFL
play-by-play data to evaluate potential strategies. We define a strategy
broadly as any set of principled decisions consistently made by an NFL
team during a game. An example, albeit possibly extreme, is for a team
to employ only passing plays rather than a mixture of passes and runs
while on offense. This is a simple strategy, but one we can nonetheless
examine using our package. To examine a particular strategy, we sample
plays satisfying the criteria of the strategy at hand. Going back to our
simplistic example, we would sample only passing plays if we wanted to
see what happened when a team only passes the football.

Sampling data to make estimates, inferences, or decisions about a larger
population is at the core of statistics and lends important rigor to our
method. In our package, we select probability samples according to a
simple random sample with replacement from our population of interest
(NFL play-by-play data) to produce unbiased and representative results.
An excellent resource for more on statistical sampling can be found in
\cite{lohr}.

The package relies on NFL play-by-play data available via the NFL's
Application Programming Interface. These datasets are accessible within
R using either the \texttt{nflscrapR} or the \texttt{nflfastR} R
packages \citep{scrapr,fastr} (or by downloading it directly from the nflscrapR-data or nflfastR-data websites \citep{yurko,baldwin}. The \texttt{NFLSimulatoR} package includes two functions, \texttt{download\_nflscrapr\_data()} and
\texttt{download\_nflfastr\_data()}, for directly downloading regular-,
pre-, or post-season NFL play-by-play data from either source for
several years, currently from 2009 - 2019. Each year contains
approximately 48,000 plays of data. In addition, we include a function
called \texttt{prep\_pbp\_data()} to eliminate extraneous information
and prepare the NFL data for use in \texttt{NFLSimulatoR} functions.

Our package is built primarily on the function \texttt{sample\_play()}.
This function samples from NFL data according to a given strategy for a
particular down and distance. The strategy is passed to the function via
the \texttt{strategy} parameter. Down and distance information refer to
what down it is (1 - 4), how many yards are required for a first down,
and the yardline at which the play occurs (1 - 99). The down is passed
to the function via the \texttt{what\_down} parameter, the distance to
go is passed via the \texttt{yards\_to\_go} parameter, and the yardline
is passed via the \texttt{yards\_from\_own\_goal} parameter. Our
sampling is done randomly and so we are confident in the outcomes from
the simulations. However, some combinations of sampling parameters
(strategy, down, distance, yardline) rarely occur in an NFL game. For
example, it may be there are few or no plays where a team had the ball
on 3rd down, on the 47th yardline, with 15 yards to go for a first down,
and chose to run the ball. In such cases we widen our sampling range to
include plays from yardlines close the to the yardline of interest or
with one less yard to go for a first down (the user can also choose a
window to expand the yardline selection via the
\texttt{window\_yards\_from\_own\_goal} parameter). We have built
flexibility into the \texttt{sample\_play()} function so the user can
seamlessly implement it in their unique settings.

The other main function of interest in the package is called
\texttt{sample\_drives()}. This allows the user to simulate a series of
plays by one team (a drive) following some specific strategy versus
another team employing a ``normal'' strategy. By ``normal'' we mean the
plays of the opposing team are simply sampled at random from all plays
without a specific strategy in mind. The \texttt{sample\_drives()}
function shows how a specific strategy is expected to perform if
implemented during an NFL game when the opposing team is employing the
status quo. The function can either sample drives until one team scores,
or it can sample a single drive and return the outcome of the drive
(i.e., touchdown, field goal, punt, or turnover). By simulating many
drives one can identify statistics such as expected points per drive and
proportion of drives resulting in a score for a variety of strategies.
The \texttt{sample\_drives()} function takes parameters for the number
of simulations to be run (\texttt{n\_sims}), the starting yardline of
the simulations (\texttt{from\_yard\_line}), the strategy
(\texttt{strategy}), and if the simulation is of a single drive
(\texttt{single\_drive}). Within \texttt{sample\_drives()}, the function
\texttt{down\_distance\_updater()} updates the down, distance, and yards
to go and then samples the next play from all plays satisfying the
updated criteria.

To demonstrate the use of this software and to offer an idea of how to
extend our work, we provide two strategies in the package. The first is
a strategy related to fourth-down decision making and the second is
associated with how often a team should pass (or run) the football.
Within the fourth down strategy we include several sub-strategies to make
a decision about going for it or not on fourth down. As mentioned above,
the fourth down strategy has been studied in the academic domain, see
e.g. \cite{yam2019lost} and \cite{romer2006firms}. We include it in this
manuscript due to its popularity and to give our own perspective on this
well-known problem. In the next section, we discuss these two strategies
in more detail.

The \texttt{NFLSimulatoR} package is available on CRAN (Comprehensive R
Archive Network) and the latest developmental version is available on
github. Adding this package to CRAN was an important step to make sure
our package passed rigorous software checks and to make installation
simpler. 
Additional package details related to issues, recent changes, etc. can
be found at the \href{http://datacolorado.com/r/NFLSimulatoR}{NFLSimulatoR
website}. The package can be installed within R using either option
given below.

\begin{Shaded}
\begin{Highlighting}[]
\CommentTok{## From CRAN}
\KeywordTok{install.packages}\NormalTok{(}\StringTok{"NFLSimulatoR"}\NormalTok{)}
\CommentTok{## From Github}
\KeywordTok{install.packages}\NormalTok{(}\StringTok{"remotes"}\NormalTok{)}
\NormalTok{remotes}\OperatorTok{::}\KeywordTok{install_github}\NormalTok{(}\StringTok{"rtelmore/NFLSimulatoR"}\NormalTok{)}
\end{Highlighting}
\end{Shaded}

\hypertarget{sec:apps}{%
\section{Applications}\label{sec:apps}}

\hypertarget{fourth-down-strategy}{%
\subsection{Fourth Down Strategy}\label{fourth-down-strategy}}

The first strategy we examine concerns fourth down decision making. This
is one of the most well-known and discussed NFL strategies. On a fourth
down the offensive team has two options: go for it or kick. If they
kick, they can either punt the ball and allow the other team to take
offensive position or kick a field goal. The other option a team has on
fourth down is to attempt to run or pass the ball and gain enough yards
for a first down. Historically, NFL coaches tend to not go for it on
fourth down unless time is running out and/or the only possible way to
win the game involves increasing the risk of a turnover for the
potential benefit of a first down. However, thanks to the analytics
movement, teams are beginning to challenge the status quo.

In 2006, Romer began the discussion about optimal decision making on
fourth down by estimating the expected point value of kicking versus
going for it on fourth down. This was done by estimating the value of a
team having the ball at each yardline on the field. These values were
estimated from NFL play-by-play data from 1998, 1999, and 2000 \citep{romer2006firms}. This work was updated in 2013 via the New York Times' Fourth Down Bot \citep{fdbot}. Burke and Quealy use a similar calculation of the value of being at each yardline and then estimate the probability of gaining enough yards for a first down. The expected points for some fourth down can be calculated as the product of the probability of
securing a first down and the point value of a first down at the
specific yardline added to the product of the probability of not
securing a fourth down and the point value of the other team taking
possession at the given yardline.

The estimated value of being at a given yardline takes into account
field goals and the expectation can be either positive or negative. If
it is positive, the Fourth Down Bot recommends going for it on fourth
down. \cite{yam2019lost} used data from the New York Times' Fourth Down Bot in a causal
analysis and determined, on average, if teams employed the (more
agressive) strategy of the Fourth Down Bot they would enjoy
approximately 0.4 more wins per year. In the NFL where there are only 16
games in a season, 0.4 is a substantial increase in wins. For further
examination into the history of fourth down decision making see \cite{yam2019lost}.

Because this strategy is of such interest we include it in our package.
We offer five sub-strategies regarding decision making on fourth downs
to compare various methods. The first is called the \emph{empirical}
sub-strategy. Here, our functions simply select the fourth down play at
random from among all similar plays (i.e., similar with respect to down,
distance and yardline). The majority of the time this will be a punt or
field goal attempt, but there are occasions where a team may try for a
fourth down (perhaps if there is very little yardage needed for a first
down and the yardline is close to the opposing endzone). The second
sub-strategy is \emph{always go for it} and samples non-kicking plays
from the given down and distance. In this sub-strategy we do not require
the sampling to be exclusively from fourth down plays. In fact, we
expand the pool of potential plays to sample from on each of downs two
through four. That is, we sample from downs \(d\) and \(d-1\) on down
\(d\), for \(d = 2, 3, 4\). We assume the impact of, and mental anxiety
among, players due to it being fourth down is negligible because the
defensive team would have similar anxieties, the players are
professional and should be more immune to such inhibitions, and because
previous literature followed this procedure (e.g., \cite{romer2006firms} used third downs instead of fourth downs) . The third sub-strategy is
\emph{never go for it} and in it the team always punts or kicks a field
goal. This offers us a conservative strategy to study, and we simply
sample kicks (and their outcome) from the given location.

The fourth sub-strategy is \emph{go for it if yardage is smaller than
Y}. Here we let the user set the parameter \(Y\) to be the value of the
yards required for a first down. If the distance for a first down is
less than or equal to \(Y\) the strategy says to go for it, and to kick
if the distance is greater than or equal to \(Y\). This allows the
examination of a stricter sub-strategy but one offering a trade-off
between \emph{always go for it} and \emph{never go for it}. This
sub-strategy is likely more palatable for NFL teams since having a rule
to go for it on fourth if there is always less than, say, 1 yard to go
for a first down might be more acceptable than always going for it. The
final sub-strategy is \emph{expected points}. Here we use the expected
points estimated from the \texttt{nflscrapR} R package to find the
expected points at each yardline on the field. We further empirically
estimate the probability of gaining a first down and making a field
goal. Then we solve for the expected value of going for it, punting it,
and kicking a field goal. The decision is made by selecting the choice
which maximizes this expected points value. This last sub-strategy is
the most analytically reliant, and best mirrors current literature.
Because we offer these sub-strategies within a free software package
they can be re-run each season as more data becomes available allowing
analysts to make recommendations which include the most recent NFL data.

We compare these sub-strategies by plotting the percent of drives
resulting in no score, a field goal, or a touchdown for the five
sub-strategies. For the \emph{go for it if yardage is smaller than Y}
option we let \emph{Y=5}. For this and subsequent fourth down analyses,
we only keep plays occuring before the final 2 minutes of each half of
the game and only plays where one team is within 28 points of the other.
This allows us to remove any plays that result from extreme decision
making because the outcome of the game is all but determined. We use
play-by-play data from both 2018 and 2019.

\begin{figure}
\centering
\includegraphics{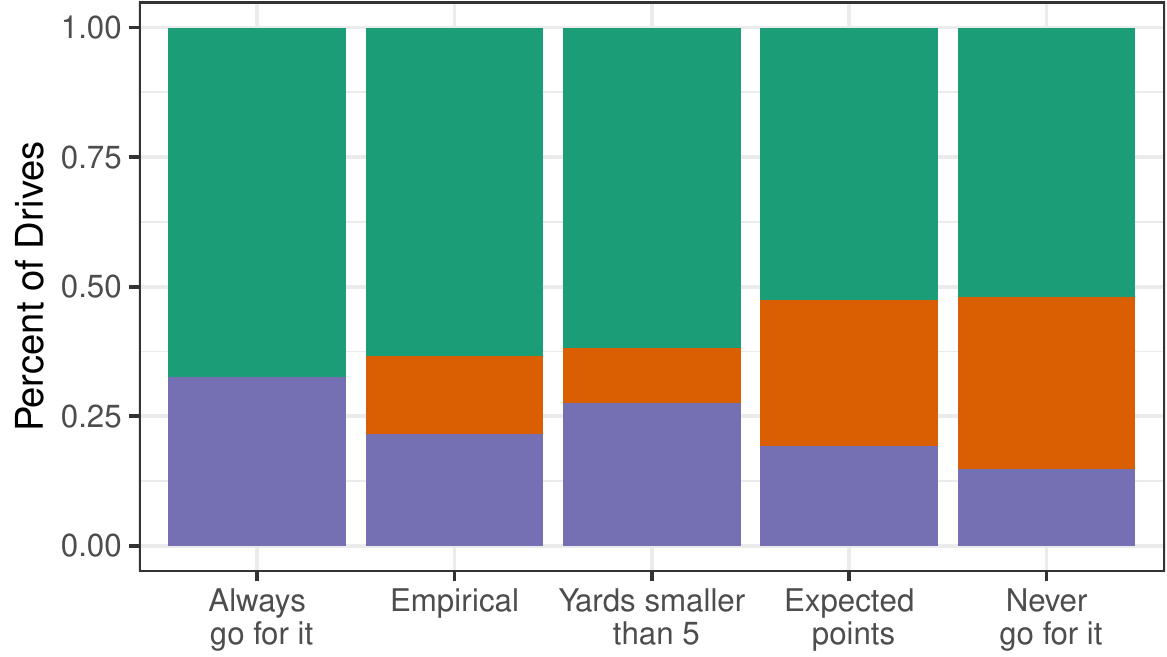}
\caption{\label{fig:fourth-down-perc-score}The percentage of simulated
drives that resulted in no score (green), a field goal (orange), or a
touchdown (purple) in 2018 and 2019, for the fourth-down sub-strategies}
\end{figure}

For the simulations, we generate 10000 drives for each sub-strategy
starting at the 25 yard line for all plays from these two regular
seasons. This corresponds to the usual starting position to begin a half
or after an opposing team scores (assuming the kickoff is a touchback).
For each drive we use the \texttt{sample\_drives()} function and set the
\texttt{single\_drive} argument equal to \texttt{TRUE}. Thus, we only
care about simulating one drive and storing its outcome for each
simulated drive. In other words, we start each drive with first down and
ten yards to go from the 25 and sample plays accordingly. The summarized
results are displayed Figure \ref{fig:fourth-down-perc-score}.

From this figure we see the \emph{never go for it} strategy offers the
largest probability for scoring on a single drive with the majority of
the scores coming from field goals. The \emph{expected points} strategy
has the second largest percentage of simulated drives resulting in a
score, followed by \emph{yardage smaller than 5 yards}, and then the
\emph{empirical} sub-strategy. For further investigation, in Table
\ref{tab:tab1} we examine the percent of drives resuling in a field goal
(FG) or touchdown (TD), the average score per drive (assuming a
touchdown always results in 7 points), and a 95\% confidence interval
for the average score, for the 5 sub-strategies.

\rowcolors{2}{gray!25}{white}
\begin{table}[H]
\begin{center}
\caption{\label{tab:tab1} A summary of 10000 simulations (2018 and 2019 data) for each fourth down sub-strategy. The percentage of drives (out of 1000) ending in either a field goal or touchdown, the average score, and 95\% confidence intervals for each sub-strategy is reported.}
\begin{tabular}{ lrrrr }
\toprule
\addlinespace[.1cm]
\multicolumn{1}{l}{\bf Sub-strategy} & \multicolumn{1}{c}{Field Goals} & \multicolumn{1}{c}{Touchdowns} & \multicolumn{1}{c}{Mean Score} & \multicolumn{1}{c}{$95\%$ CI} \\
\midrule
{\em Always Go} & 0\% & 33\% & 2.28 & (2.22, 2.35) \\
\addlinespace[.1cm]
{\em Empirical} & 15\% & 22\% & 1.96 & (1.91, 2.02)\\
\addlinespace[.1cm]
{\em Expected Points} & 28\% & 19\% & 2.19 & (2.14, 2.25) \\
\addlinespace[.1cm]
{\em Never Go} & 33\% & 15\% & 2.03 & (1.98, 2.08)\\
\addlinespace[.1cm]
{\em Go if Yards $<$ 5} & 11\% & 27\% & 2.24 & (2.18, 2.30) \\
\toprule
\end{tabular}
\end{center}
\end{table}
 
From Table \ref{tab:tab1}, the fourth down sub-strategy with the largest
average points per simulated drive is \emph{always go for it} (average
of 2.28 points) followed by \emph{yardage smaller than 5 yards} (average
of 2.24 points), and \emph{expected points} (average of 2.19 points). We
also see the \emph{always go for it} sub-strategy is boom or bust
resulting in only touchdowns or no scores. Interestingly the
\emph{yardage smaller than 5 yards} has an average score similar to
\emph{always go for it}, yet it does recommend field goals to be taken.
The confidence interval for the \emph{yardage smaller than 5 yards} mean
score is also narrower than that for \emph{always go for it}. Taking
this into account along with the fact that the averages of these two
sub-strategies are so close, a recommendation for a team nervous about
always going for it on fourth down might be to always go for it if there
are less than five yards to go for a first down, regardless of field
position. Figure \ref{fig:fourth-down-perc-score} shows this strategy
will produce scoring drives more often and has nearly the highest
average score per drive.

If a team wishes to pursue this sub-strategy (going for it on fourth if
the yards to go is less than five yards) a logical next question is:
what about other \emph{yards to go} values? That is, what if the team
went for it if the yards required for a first down are 4, or 6, or
something else entirely? Figure 2.1 shows the percent of drives
resulting in a score for a range of \(Y\) values. Figure 2.2 displays
the average (and 95\% confidence interval) score per drive for the
various \(Y\) values, and Figure 2.3 gives the average (and 95\%
confidence interval) yardline at which the ball is turned over when the
drive does not result in a score.
\clearpage
\begin{figure}
\centering
\includegraphics{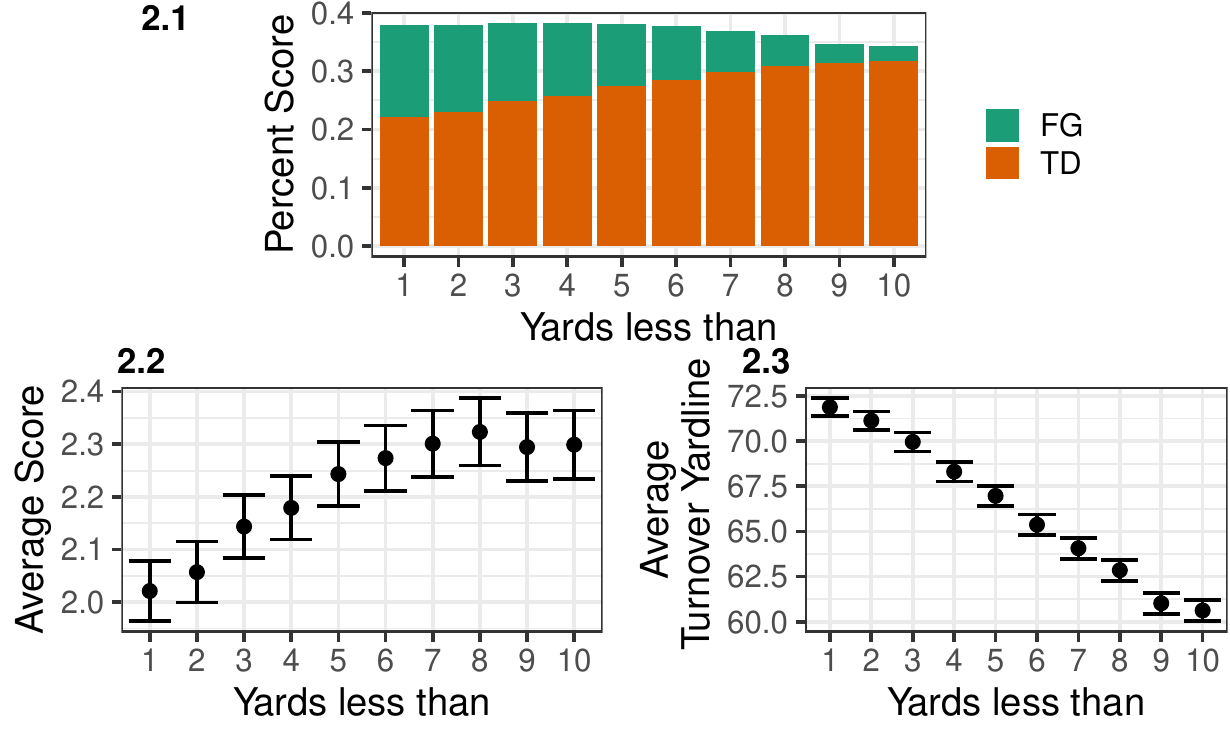}
\caption{\label{fig:yds-less-than} 2.1: The percentage of simulated
drives that resulted in a field goal (green) or a touchdown (orange);
2.2: Average score per drive for the \emph{yardage less than Y yards}
sub-strategy as a function of \(Y\); 2.3: Average turnover yardline
resulting from the \emph{yardage less than Y yards} sub-strategy as a
function of \(Y\)}
\end{figure}

In Figure 2.1 the largest percent score value (of about 38\%) is nearly
exactly achieved by \(Y\) values of 3, 4, and 5. Figure 2.2 shows the
\(Y\) values of 8 has the top average score per drive values, and this
average decreases as \(Y\) decreases. Figure 2.3 shows the average
turnover yardline gets further away from the offensive teams goal for
larger values of \(Y\). Taking all this together, a value of 5 yards may
be the best option for the fourth down sub-strategy \emph{go for it if
yardage less than \(Y\)} because it has nearly the highest percent score
value, a higher average score than all smaller \(Y\) values, a more
advantageous average turnover yardline than all larger \(Y\) values, and
(speculatively) may be more acceptable by NFL coaching staffs than a
value of, say, \(Y = 8\).

Here, we caution the reader that this is by no means a causal
investigation of fourth down strategies. Indeed, we could further
analyze the data by evaluating the performance of a specific
sub-strategy amongst better or worse teams, but do not do so as our
primary purpose is to demonstrate the usefulness of the
\texttt{NFLSimulatoR} package and its core functionality.

\hypertarget{runpass-percentage}{%
\subsection{Run/Pass Percentage}\label{runpass-percentage}}

\cite{levitt09} and \cite{hermsmeyer} argue NFL teams should
pass more often. In this section we investigate this thesis using the
simulation-based approach of \texttt{NFLSimulatoR}. Though perhaps
simple on its surface, examining a strategy having to do with the
proportion of plays that are a pass instead of a run proves interesting.
Even if the NFL is not as analytically forward as other professional
sports leagues, the league seems to be trending towards passing more.
The \texttt{NFLSimulatoR} package includes a strategy allowing the user
to study the effect of passing the ball more or less often.

When employing this strategy in the \texttt{sample\_play()} or
\texttt{sample\_drives()} functions, the argument \texttt{p} must be included
as a parameter. \texttt{p} is the probability a given offensive play on
first, second, or third down is a pass. To keep the strategy
straightforward, we follow an empirical procedure when the play to be
sampled is a fourth down. That is, when a fourth down situation arises
in the sample, we assume the play is simply sampled from all fourth down
plays at the given yardline (or within a neighborhood of the yardline)
and distance to go until a first down. Fourth down plays sampled at
their regular rates usually result in a punt or a field goal attempt. By
varying \texttt{p} we can study how pass proportion affects statistics such
as the expected points per drive, the proportion of drives resulting in
a score, among a host of other metrics.

\begin{figure}
\centering
\includegraphics{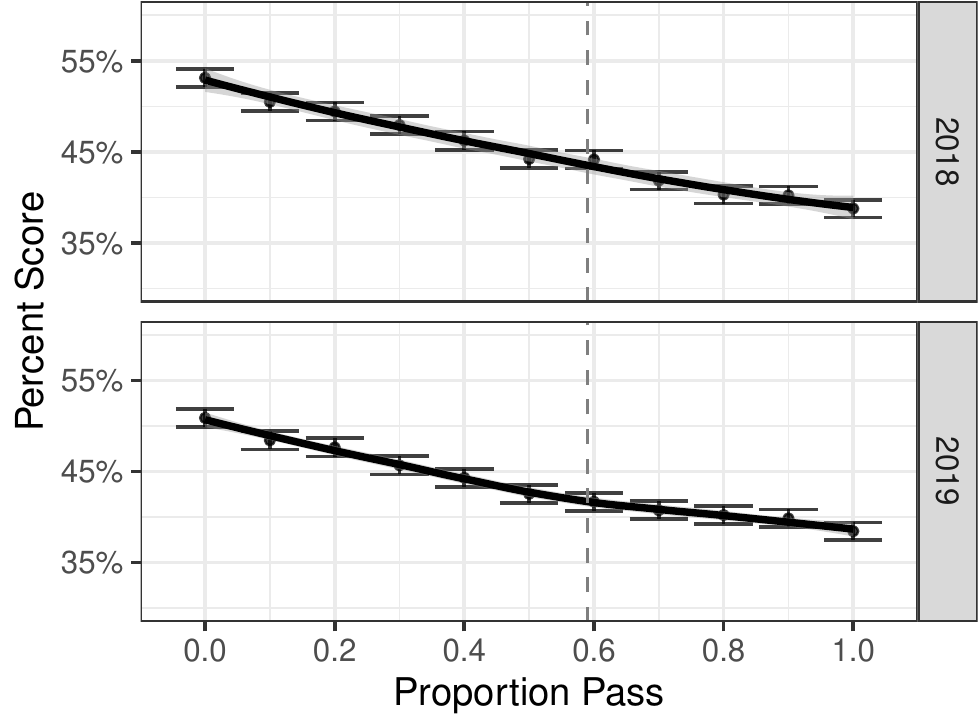}
\caption{\label{fig:pass-rush-all-facet}The percentage of simulated
drives that resulted in a score (touchdown or field goal) in 2018 and
2019. The dashed line represents the actual proportion of passing plays
on first, second, and third downs in both years.}
\end{figure}

\begin{figure}
\centering
\includegraphics{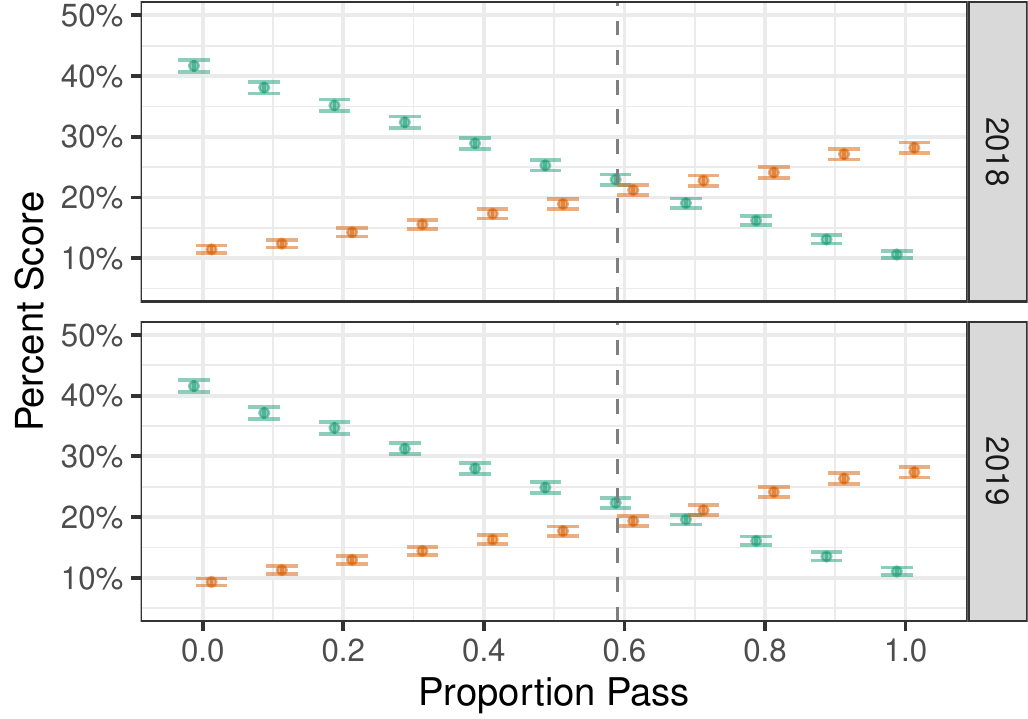}
\caption{\label{fig:pass-rush-by-type}The percentage of simulated drives
that resulted in either a touchdown (orange) or a field goal (green) in
2018 and 2019. The dashed line represents the actual proportion of
passing plays on first, second, and third downs in both years.}
\end{figure}

Figure \ref{fig:pass-rush-all-facet} shows the proportion of simulated
drives resulting in a score for the offensive team (field goal or
touchdown) in 2018 and 2019. Note that we include a vertical dashed line
showing the league-wide proportion of passing plays on first through
third downs. This proportion of passing (running) plays on first through
third downs was roughly 59\% (41\%) in both 2018 and 2019. At first
inspection this figure suggests passing more often results in scoring
\textbf{less} on average. Obviously this initial glance requires more
scrutiny and indeed, subsetting by the type of score reveals additional
insight. Specifically, Figure \ref{fig:pass-rush-by-type} shows the same
data subsetted by the type of score: either a touchdown or field goal.
There is a clear trend showing more touchdowns are scored as the
proportion of plays that are passes increases.

\begin{figure}
\centering
\includegraphics{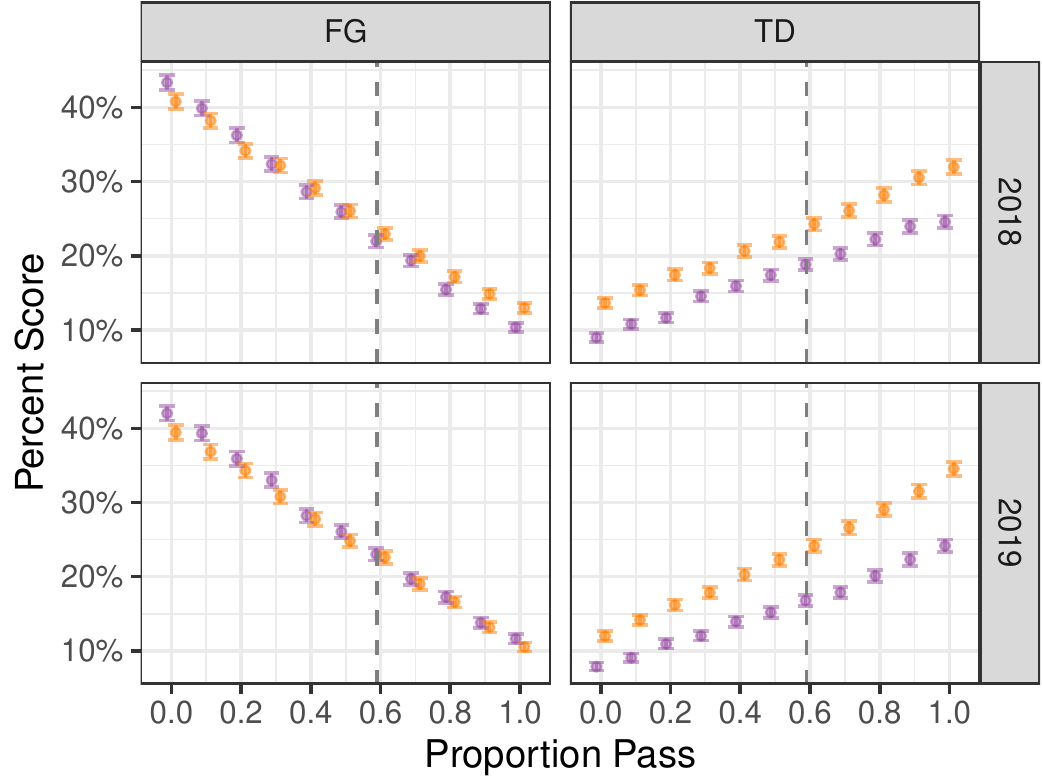}
\caption{\label{fig:pass-rush-facet} The percentage of simulated drives
that resulted in a score by type (touchdown or field goal) in 2018 and
2019 colored by playoff teams (orange) versus non-playoff teams
(purple).}
\end{figure}

Next, we look at the percentage of drives resulting in a score broken
down by the quality of the team. In this case we subset by whether or
not a team made the playoffs, and use playoff appearance as a proxy for
quality. To do this we simulate one set of drives by sampling plays from teams that made the playoffs and another set of drives by sampling teams that did not.
Figure \ref{fig:pass-rush-facet} shows drives using plays from the
better teams (i.e., playoff teams) tend to result in a score more often
when employing a heavier passing-based strategy than the drives from
non-playoff teams.

\begin{figure}
\centering
\includegraphics{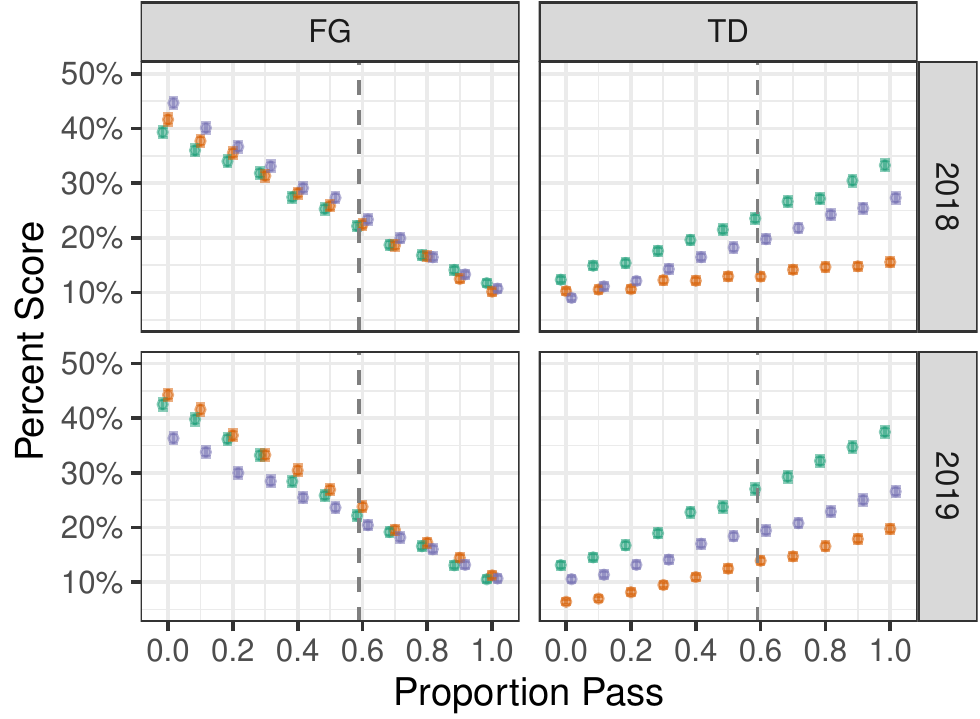}
\caption{\label{fig:pass-rush-qbr}The percentage of simulated drives
that resulted in a score by type (touchdown or field goal) in 2018 and
2019 colored by overall team passer rating classification: High (green),
Medium (purple), and Low (orange).}
\end{figure}

Again, we stress that our approach is not causal in any sense of the
imagination. That is, we are not saying that passing more will
necessarily lead to more scores, particularly if the team has a
sub-standard quarterback. This result, of course, is likely confounded
by playoff teams (traditionally) having better quarterbacks. Thus, we
next subset the pool of plays by each team's overall passer rating (RTG)
and sample plays from three distinct pools: High, Medium, and Low passer
rating teams. A team in the pool of High passer rating teams had an
overall rating falling into the upper-most tercile of teams. The pools
for Medium and Low are similarly defined. The results of the simulated
drives using these groups are displayed in Figure
\ref{fig:pass-rush-qbr}. Here, we see the teams in the upper tercile of
passing ratings score more touchdowns as the proportion of passing plays
increases than teams in the other two groups. However, the percent field
goals scored as a function of the proportion of passing plays is similar
for all the three team groupings.

Our overall conclusion, based on these simulations, is that passing more
should lead to a higher percentage of touchdowns scored. This conclusion
is not uniformly true across all types of teams, however. That is, the
better teams, or those teams with a higher-quality quarterback relative
to the rest of the teams in the league, will benefit much more than the
others. 

Finally, we should mention that the time remaining in a game, as well as 
other variables, could confound these results as well. That is, teams that
are winning in the fourth quarter might elect to rush more often in order 
to ``shorten the game'' and losing teams might pass more. Therefore, a future study might include a time-in-game parameter, $\tau$, in order to sample from plays within a window around $\tau$. 

\hypertarget{sec:conc}{%
\section{Conclusions and Discussion}\label{sec:conc}}

Even though the NFL has existed since 1920, teams are still seeking
inefficencies in the game to exploit. There always seems to be a
brilliant new coach ready to introduce a new strategy to push teams to
more success. The purpose of creating \texttt{NFLSimulatoR} is to give
the wider community a tool to examine a multitude of NFL strategies. The
package contains a set of robust and statistically sound tools to
simulate plays and drives to examine NFL game plans. This package will
also age well as it can be continually updated with data from the most
recent NFL season. Another use for the package is to examine the
ramifications of rule changes by the league. This would allow the league
to take a data-driven approach to such changes. One example of a rule
change that has been debated is eliminating the kickoff after a score.

In the package we include two strategies of interest, passing versus
rushing the ball and going for it or not on fourth down. We have
examined each strategy in this paper as examples of possibilities for
the package. We imagine many extensions of our work, including
strategies regarding whether to run or throw on first down, what play
works best after a penalty or timeout, and what plays to run in the
first or last few minutes of a game or quarter. Another obvious extension to this work is the implementation of more game- or team-specific scenarios. For example, we might study game-specific strategies by sampling from plays satisfying, or nearly satisfying, additional simulated in-game characteristics such as time-in-game, current score, weather, location, among a host of additional parameters. Taking this concept a step further, we could also assign higher sampling probabilities to plays most closely matching these given in-game characteristics.

In addition, we might study team-specific strategies if we assume some teams perform better at various aspects of the game than others. For example, a team could have an excellent run game but a poor pass game, or excel at throwing the ball fewer than 20 yards but struggle when the throw is over 20 yards. In such cases, we could only sample plays which match a given team’s profile to study the outcomes of strategies a specific team is more likely to employ. Our methodology is robust enough to include these subsetting parameters in any of the sampling functions given in \texttt{NFLSimulatoR}. Furthermore, the simulations can either be at the individual drive level (as we did) or by evaluating a strategy over the course of an entire game. 

Another possible extension of our work is to choose strategies at random to implement. By randomly selecting and setting various parameters for a strategy (or combination of strategies) one can compare a bevy of strategies. As the adoption of \texttt{NFLSimulatoR} grows and accumulates more available strategies, randomly choosing and combining strategies and then testing them seems immensely useful. Harnessing computational power to examine a plethora of strategies (such as: always go for it on fourth down while also passing on 77\% of first down plays) will only lead to further optimization of in-game decision making in the NFL.

We welcome collaboration from the sports analytics community and hope
for contributions to our package, which are easy to make given its
open-source nature. The fact that there are so many possible analysis options of game strategies makes us more excited about the existence of the \texttt{NFLSimulatoR} package because now the wider sports analytics community can take our initial work and extend it. As an example, recently a new model-based fourth down decision maker was introduced by Ben Baldwin, an author of the previously mentioned \texttt{nflfastR} package \cite{baldwin_athletic}. This is exactly the sort of contribution we hope will be added to the \texttt{NFLSimulatoR} package. Such a strategy could be integrated and tested within the simulation based framework we created and shared with the community at large. We look forward to what new strategies will be devised and tested and hope to see even more analytics used in the NFL and other sports leagues.

Finally, we stress that \texttt{NFLSimulatoR} is in its infancy with the current release being v0.3.1. As previously mentioned, we encourage interested parties to contribute to the package as this project evolves. Contributions could be in the form of new strategies, vignettes showing new and interesting analyses, or simply code enhancements. For reproducibility we have included code to generate the figures and table used in this paper as a github repository located at \href{https://github.com/williamsbenjamin/nflsimulator_aoor}{\text{github.com/williamsbenjamin/nflsimulator\_aoor}}. 

\bibliography{bibliography.bib}
\bibliographystyle{spbasic}
\end{document}